\documentclass[aps,twocolumn,showpacs,preprintnumbers,amsmath,amssymb]{revtex4}
\usepackage{graphicx}
\usepackage{dcolumn}
\usepackage{bm}

\begin{document}


\title{Higher Order Terms of Kerr Parameter for Blandford-Znajek Monopole Solution}

\author{Kentarou Tanabe}
\email{tanabe@yukawa.kyoto-u.ac.jp}
\author{Shigehiro Nagataki}
\email{nagataki@yukawa.kyoto-u.ac.jp}
 
\affiliation{Yukawa Institute for Theoretical Physics, Kyoto
University, Oiwake-cho Kitashirakawa Sakyo-ku, Kyoto 606-8502, Japan}

\date{\today}

\begin{abstract}
Blandford-Znajek mechanism, by which the rotational energy of a black
hole is extracted through electromagnetic fields, 
is one of the promising candidates as an essential
process of the central engine of active compact objects such as
Gamma-Ray Bursts.
The only known analytical solution of this mechanism is the
perturbative monopole solution for 
Kerr parameter $a$ up to the second order terms. In order to apply Blandford-Znajek mechanism to 
rapidly rotating black holes, we try to obtain the perturbation solution 
up to the fourth order. 
As a result, we find that the fourth order
terms of the vector potential diverge at infinity, which implies that
the perturbation approach breaks down at large distance from the black
hole. Although there are some uncertainties 
about the solution due to the unknown boundary condition at infinity for the fourth order terms, 
we can derive the evaluation of the total energy flux extracted from the black hole
up to fourth order of $a$ without any ambiguity. 
Further more, from the comparison between the numerical solution that
is valid for $0<a<1$ and the fourth order solution, we find 
that the fourth order solution reproduces the numerical result better
than the second order solution.
At the same time, since the fourth order solution does not match well
with numerical result at large $a$, we conclude that more higher order
terms are required to reproduce the numerical result.  
\end{abstract}

\maketitle

\section{introduction}
Gamma ray bursts(GRBs) are the most powerful explosions in the
Universe~\cite{meszaros06,piran05}. GRBs release
$10^{51}-10^{53}\text{erg}$ in a few seconds, which corresponds to a
rest mass energy of the sun. 
The mechanism of the central engine of GRBs is not still understood well.
However, from the fact that some GRBs are associated with
supernova explosions~\cite{galama98,hjorth03}, we can guess that there might be a 
black hole at the center of a GRB and its energy source may be the
rotation energy of the black hole.
In order to extract the rotational energy of the black hole, the
Blandford-Znajek mechanism was proposed~\cite{bz77,mckinney04,lee00}. 
In this mechanism, the rotational energy is extracted in a form of the
Poynting flux that is realized by the surrounding electromagnetic fields. 
The scale of extracted
energy from the black holes is, in a rough estimate~\cite{thorne86}, 
\begin{equation}
 \begin{split}
 \dot{E}&\sim B^2\Omega_{H}^2M^2 \\
             &\sim 10^{51}\text{erg/s}\left(\frac{a}{M}\right)^2
 \left(\frac{M}{10M_{\odot }}\right)^2
 \left(\frac{B}{10^{15}\text{G}}\right)^2 
  \end{split}
\end{equation}
where $B$ is a magnetic field, $M$ is the mass of the black hole, $a$ is
the specific angular momentum of black hole (Kerr parameter). 
Compared with the energy scale of GRBs, this energy scale seems to be
a good one.

Although Blandford and Znajek (1977) obtained an analytical solution for this mechanism 
by assuming the shape of the magnetic field (monopole magnetic field), this solution 
has two points that should be investigated further. 
First, this solution is the perturbative solution for 
Kerr parameter $a$. Thus, this solution is valid only for a slowly
rotating black hole ($a\ll M$). 
Second, it is very difficult to impose proper boundary conditions. 
The equation we have to solve is the second order differential
equation, thus we must impose two boundary conditions  
for a solution. 
In the original paper, assuming force free condition in all region, 
they imposed boundary conditions such that at event horizon the magnetic field is 
finite and at infinity the solution should connect with another force free solution 
obtained in Newton gravity~\cite{michel73}. The boundary condition at event horizon is 
plausible~\cite{thorne86}, but since at large $r$, there is no guarantee that we can 
use force free condition, we have to be careful for the boundary condition at infinity.

The aim of this paper is to evaluate more accurately the energy flux
extracted from a rotating black hole by the Blandford-Znajek
mechanism, i.e. to derive the fourth order terms of Kerr parameters
for the monopole solution. 
Although there is two natural boundary conditions for the second
order solution, proper boundary conditions are not known
for the higher order solutions.
Thus, when we solve the field
equations, we impose a boundary condition only at the event horizon
while we do not impose any boundary condition at large $r$ in this
study, and we investigate the behavior of the magnetic field at large
$r$. In particular, we examine whether we can impose the force free
condition at infinity in order to check the validity of the force free
boundary condition.
Further, we compare the analytical solution we obtain in this study
with the numerical calculation in order to evaluate the contribution
of the fourth order terms to the extracted energy flux from the black hole.

The plan of this paper is as follows. In section $2$, we solve the
perturbation equations for Kerr parameter $a$ by imposing a boundary
condition at the event horizon and estimate the total energy flux extracted
from a rotating black hole. 
Here, we use Kerr-Schild coordinate since this coordinate is regular at the
event horizon. Although the second order solution is shown to be
regular in a hole region, we show the fourth order solution diverges
irrespective of any boundary condition at infinity, which means the
perturbation approach breaks down at large distance from the black hole.
Our notation follows McKinney and Gammie~\cite{mckinney04} in section
$2$. In section $3$, we
compare the analytical solution obtained in section $2$ with numerical
solution which is valid for $0<a<M$. In section $4$, we summarize and
discuss our results.

\section{method of calculation}

As stated in section $1$, in Blandford-Znajek mechanism, rotational
energy of a black hole is extracted through electromagnetic fields
and this extracted energy may be the energy source of GRBs. 
When the energy is extracted through electromagnetic fields, 
it propagates in the form of the Poynting flux vector. 
Here, we give the formula of Poynting flux vector in Kerr
space-time for arbitrary shape of magnetic fields. 
As mentioned above, we use Kerr-Schild coordinate here for convenience:  
\begin{align}
ds^2 &= (-1+\frac{2r}{\Sigma})dt^2 + (1+\frac{2r}{\Sigma})dr^2 + \Sigma d\theta ^2 \notag \\
& + \frac{\sin ^2\theta}{\Sigma} \left((r^2+a^2)^2-\triangle a^2\sin
^2\theta \right) d\phi ^2  + \frac{4r}{\Sigma}dtdr \notag \\ 
& - \frac{4ar\sin ^2\theta}{\Sigma}dtd\phi ,- 2a\sin ^2\theta \left(1+\frac{2r}{\Sigma}\right) drd\phi
\end{align}
where
\begin{gather}
\Sigma = r^2 + a^2\cos^2\theta \\
\triangle = r^2 - 2r + a^2
\end{gather}
and we use the unit in which $G=c=M(\text{black hole mass})=1$. The energy momentum tensor 
of electromagnetic fields is written in terms of $F_{\mu\nu}=\partial_{\mu}A_{\nu}-\partial_{\nu}A_{\mu}$ 
\begin{equation}
T^{EM}_{\mu\nu} = F_{\mu\rho}F_{\nu}^{\,\,\rho} - \frac{1}{4}F_{\alpha\beta}F^{\alpha\beta}g_{\mu\nu}.
\end{equation}
$-T^{r}_{\,\,t}$ component of $T^{\mu}_{\,\,\nu}$ is the $r$ component of Poynting flux vector.

We can expect that matter's contribution to the total energy can be
ignored near the black hole, 
so force free condition is valid~\cite{bz77}. Force free condition is represented by electromagnetic 
tensor $F_{\mu\nu}$ as follows:
\begin{equation}
F_{\mu\nu}J^{\nu} = 0 \label{eq:ff}
\end{equation}
where $J^{\mu}$ is the current vector. This approximation means that
the inertial force by matter is ignored and 
the total energy momentum tensor is composed of electromagnetic one only:
\begin{equation}
T^{\mu\nu}_{total} = T^{\mu\nu}_{matter} + T^{\mu\nu}_{EM} \cong  T^{\mu\nu}_{EM}.
\end{equation}
In other word, matter energy is ignorable compared with electromagnetic energy. Thus, 
in the force free region, equation of motion of electromagnetic field
becomes as follows:
\begin{equation}
\nabla_{\mu}T^{\mu\nu}_{EM} \label{eq:eom} = 0.
\end{equation}
We use $T^{\mu\nu}$ as $T^{\mu\nu}_{EM}$ below.

Generally, the number of independent variables is six; electric fields and magnetic fields. 
However, when we impose the force free condition, there are four independent variables. 
Further, assuming the the force free condition (\ref{eq:ff}), stationarity,
and axisymmetry, there is a relation among vector potential terms
$A_{\mu}$:
\begin{equation}
A_{\phi,\theta}A_{t,r} - A_{t,\theta}A_{\phi ,r} = 0 \label{eq:ang-v}
\end{equation}
which makes the number of independent variables three.
This relation implies that part of the electric field components are represented by
magnetic field components. 
Also, we can define the angular velocity $\Omega_{F}$ of electromagnetic fields 
from (\ref{eq:ang-v}):
\begin{equation}
\frac{A_{t,\theta}}{A_{\phi ,\theta}} = \frac{A_{t,r}}{A_{\phi ,r}} \equiv \Omega_{F}.
\end{equation}
All components of the electric fields are represented by this angular velocity $\Omega_{F}$ 
and magnetic field components $B^{i} \,(i=r,\theta,\phi)$. 
After all, independent variables are $\phi$ 
component of magnetic field $B^{\phi}$, $\phi$ component of vector potential $A_{\phi}$ 
and angular velocity of electromagnetic field $\Omega_{F}$. Here, we define magnetic field 
components as follows:
\begin{gather}
B^{r} = \frac{1}{\sqrt{-g}}F_{\theta\phi}, \\
B^{\theta} = \frac{1}{\sqrt{-g}}F_{\phi r}, \\
B^{\phi} = \frac{1}{\sqrt{-g}}F_{r\phi}.
\end{gather}
Using these variables, we can evaluate the energy flux
$-T^{r}_{\,\,t}$ of electromagnetic field as
\begin{equation}
-T^{r}_{\,\,t} = -2(B^{r})^2\Omega_{F}\left(\Omega_{F} - \frac{a}{2r}\right)\sin^2\theta - B^rB^\phi\Omega_{F}\triangle\sin^2\theta.
\end{equation}
On the event horizon $(r=r_{+} \Rightarrow \triangle=0)$, this formula becomes 
\begin{equation}
-T^{r}_{\,\,t} = 2(B^r)^2\Omega_{F}r_{+}\left(\Omega_{H} - \Omega_{F}\right)
\end{equation}
where $\Omega_{H}$ is the angular velocity of the black hole. Thus rotational energy can be extracted 
through electromagnetic fields in the force free region if and only if $0<\Omega_{F}<\Omega_{H}$.

What we want to evaluate is the total energy flux of the extracted
energy from the black hole. 
Its total energy flux $\dot{E}$ is represented as the integration of
the energy flux at $r=\text{const.}$ surface. 
\begin{equation}
\dot{E} =-2\pi\int_{0}^{\pi}\sqrt{-g}T^{r}_{\,\,t}d\theta \label{eq:ef}
\end{equation} 
From stationarity and axisymmetry of Kerr space time, this total energy flux $\dot{E}$ 
is independent of $r$ at which the integration is operated.

The formula (\ref{eq:ef}) which we derive here can be used for any shape of magnetic fields. 
In this section we evaluate this total energy flux for the monopole magnetic field. Since it is 
very difficult to solve the equation of motion (\ref{eq:eom})
analytically, we treat Kerr metric as the perturbation
in Kerr parameter from Schwarzschild metric to obtain analytical solutions.

\subsection{second order terms}
Here, we explain the monopole solution which Blandford and Znajek obtained by perturbation method 
for Kerr parameter $a$. At first, the unperturbed solution is the
monopole solution in Schwarzschild space time. 
This unperturbed solution satisfies 
\begin{equation}
\nabla_{\mu}^{(0)}T^{\mu}_{\,\,\nu} = 0
\end{equation}
where $(0)$ means that Schwarzschild metric is used. The first order monopole solution is
\begin{gather}
A_{\phi}^{(0)} = -C\cos\theta, \label{eq18}\\
B^{\phi (0)} = 0,  \\
\Omega_{F}^{(0)} = 0
\end{gather}
where $C$, strength of the magnetic field, is the constant. In this solution, since electromagnetic fields 
do not rotate, rotational energy of black hole cannot be extracted.

In order to obtain the monopole solution around a slowly rotating black hole, we consider 
perturbation for Kerr parameter $a$. Taking account of axisymmetry, we expand the solution up to the 
second order as follows:
\begin{gather}
A_{\phi} = -C\cos\theta + a^2 A_{\phi}^{(2)}, \\
B^{\phi} = aB^{\phi(1)}, \\
\Omega_{F} = a\omega^{(1)}.
\end{gather}
We also use the expanded Kerr-Schild metric up to the second order in $a$.

The equation we solve is $\nabla_{\mu}T^{\mu}_{\,\,\nu} = 0$. The $t,\phi$ components of this equation 
represent the conservation equation for energy and angular
momentum of the electromagnetic fields, respectively. 
The $r,\theta$ components are the trans-field equations. Here, we impose boundary conditions. 
\begin{gather}
A_{\phi}^{(2)}|_{\text{on the event horizon}} = \text{finite},  \label{eq:bc1}\\
A_{\phi}^{(2)} \rightarrow O\left(\frac{1}{r}\right) (r \rightarrow \infty ), \label{eq:bc2}\\
B^{\phi(1)}|_{\text{on the event horizon}} = \text{finite}. \label{eq:bc3}
\end{gather}
The conditions (\ref{eq:bc1}),(\ref{eq:bc3}) in Kerr-Schild coordinate corresponds to the condition 
in Boyer-Lindquist coordinate such that the magnetic field that FIDO
(fiducial observer) feels should be 
finite \cite{znajek77}, and (\ref{eq:bc2}) is required to connect the solution to the Michel's solution. 
In ref~\cite{mckinney04}, regularity at infinity and separability of
$A_{\phi}^{(2)}$ are taken as the boundary conditions. 
Although the boundary conditions are different from ours, the result
we obtain below is the same with McKinney 
and Gammie~\cite{mckinney04}. As we see in fourth order calculation, $A_{\phi}^{(4)}$ is not 
separable. Thus we do not adopt the separability as a boundary
condition in this study.

Under these boundary conditions, the solutions are derived as follows:
\begin{gather}
A_{\phi}^{(2)} = Cf(r)\cos\theta \sin^2\theta, \\
B^{\phi(1)} = -\frac{C}{4r^2}\left(\frac{1}{2} + \frac{2}{r}\right), \\
\omega^{(1)} = \frac{1}{8}
\end{gather}
where 
\begin{align}
f(r) =& \left(Li_{2}(\frac{2}{r}) -\ln(1-\frac{2}{r})\ln(\frac{r}{2}) \right)\frac{r^2(2r-3)}{8} \notag\\
&+\frac{1+3r-6r^2}{12}\ln(\frac{r}{2})+\frac{11}{72}+\frac{1}{3r}+\frac{r}{2}-\frac{r^2}{2}
\end{align}
and
\begin{equation}
Li_{2}(x) = -\int_{0}^{1}\frac{\ln (1-tx)}{t}dt. 
\end{equation}
By choosing $\omega^{(1)}=1/8$, we can use the force free condition at large $r$ consistently. 
We evaluate the total energy flux of the electromagnetic field.
\begin{align}
\dot{E} &=-2\pi\int_{0}^{\pi}\sqrt{-g}T^{r}_{\,\,t}d\theta \notag \\
           &=\frac{\pi}{24}a^2C^2. \label{eq:tef1}
\end{align}
Eq.~(\ref{eq:tef1}) is the result that Blandford and Znajek (1977) obtained.

\subsection{fourth order terms}

Here, we try to obtain the fourth order terms for Kerr
parameter. Although the method to derive the fourth order terms 
is basically same as the second order case, we have a problem with
the boundary condition as mentioned in section $1$. 
We must impose two boundary conditions. One is the
condition at the event horizon. This is same as the second order
solution. As for another boundary condition, we cannot guess anything
since we do not know where the force free condition breaks down and
which solution we can connect to. 
That is why we impose a boundary condition only at the event horizon and 
we solve the fourth order equations. We also 
examine whether we can put a force free boundary condition at large
$r$ using the derived fourth order solution.

The equation we solve is the fourth order one for Kerr parameter 
$a$ of $\nabla_{\mu}T^{\mu}_{\,\,\nu}=0$. We expand the variables as follows:
\begin{gather}
A_{\phi}=-C\cos\theta + a^{2}A_{\phi}^{(2)} + a^4A_{\phi}^{(4)}, \\
B^{\phi} = aB^{\phi(1)} + a^3B^{\phi(3)}, \\
\Omega_{F} = a\omega^{(1)} + a^3\omega^{(3)}.
\end{gather}
By checking the fourth order terms of the $t$ and $\theta$ for the
equation $\nabla_{\mu}T^{\mu}_{\,\,\nu}=0$, one can guess that the
$\theta$ dependence should be as follows:  
\begin{gather}
A_{\phi}^{(4)} = h_{1}(r)\cos\theta + h_{3}(r)\cos^3\theta + h_{5}(r)\cos^5\theta, \\
B^{\phi(3)} = g_{0}(r) + g_{2}(r)\cos 2\theta, \\
\omega ^{(3)} = b_{0} + b_{2}\cos 2\theta
\end{gather}
where $b_{0},b_{2}$ are constants. When we impose boundary condition at the event horizon 
like the second order solution, we can solve the $\phi$ component of
the equation for $g_{0},g_{2}$.
The solutions are as follows: 
\begin{widetext}
\begin{align}
g_{0}(r) = &\frac{1}{576r^5(r-2)}\left( -288+456r+92r^2+109r^3+36r^6-1152b_{0}r^3+\frac{r^4}{2}(6\pi^2 -521+1152b_{0})\right. \notag \\  
             &\left.+3r^2(8+22r-56r^2+6r^3+12r^4)+9r^4(12-11r-r^2+2r^3)(\ln (\frac{r-2}{r})\ln ( \frac{r}{2} )-Li_{2}(\frac{2}{r}))\right), \\
g_{2}(r) = &\frac{1}{192r^5(r-2)}\left( -96+120r+20r^2+125r^3+24r^5+12r^6-384b_{2}r^3+\frac{r^4}{2}(384b_{2}-95-2\pi^2 ) \right. \notag\\
             &\left.
             +r^2(24+66r-164r^2+30r^3+12r^4)+3r^4(36-33r+3r^2+2r^3)(\ln
             (\frac{r-2}{r})\ln ( \frac{r}{2}
             )-Li_{2}(\frac{2}{r}))\right).  
\end{align}
\end{widetext}
The behavior of these solutions at large $r$ are
\begin{gather}
g_{0}(r) = \frac{1152b_{0}+6\pi^2 -139}{1152r^2} +
O\left(\frac{1}{r^3}\right),  \\
g_{2}(r) = \frac{1152b_{2}-6\pi^2 +67}{1152r^2} + O\left(\frac{1}{r^3}\right).
\end{gather}
These solutions depend on the constants in the angular velocity of the
electromagnetic fields $b_{0},b_{2}$, which must be determined by
another boundary condition that we do not know. 
The total energy flux of electromagnetic field is 
\begin{align}
\dot{E}& = -2\pi\int_{0}^{2\pi}\sqrt{-g}\,T^{r}_{\,\,t} d\theta  \notag \\
         & =\frac{\pi}{24}a^2C^2 + \frac{\pi(56-3\pi^2)}{1080}a^4C^2. \label{eq:ef4}
\end{align} 
This is the value we want to evaluate. Fortunately, this result
does not depend on the undetermined constants $b_{0},b_{2}$.
Probably, it is by chance. 
On the other hand, at large $r$, we can solve the $r$ component of the
fourth order equation 
of $\nabla_{\mu}T^{\mu}_{\,\,\nu}=0$ for $A_{\phi}^{(4)}$. 
The behavior is 
\begin{equation}
A_{\phi}^{(4)} \rightarrow O(r^2)
\end{equation}
No matter how we choose $b_{0},b_{2}$ which must be determined by
a boundary condition, we cannot make $A_{\phi}^{(4)}$ $O(1/r)$ as the second
solution at large $r$. This result implies that this solution cannot
work well at large $r$ because perturbation method breaks down there. 
However, the evaluation of the total energy flux is valid because, as mentioned
above, $\dot{E}$ is independent of $r$ whether we use perturbation
method or not.

\section{comparison with numerical calculation}

In order to see how well the forth-order term works to describe the
total energy flux for large Kerr parameter, we have performed
numerical simulations of the monopole solution and compared the
results with the analytical, second-order and forth-order solutions.

We have developed a two-dimensional General Relativistic Magneto-Hydro Dynamics
(GRMHD) code following~\cite{gammie03}~\cite{noble06}.
We have adopted a conservative, shock-capturing scheme with Harten,
Lax, and van Leer (HLL) flux term~\cite{harten83} with
flux-interpolated constrained transport technique~\cite{toth00}. 
We use a third-order TVD Runge-Kutta method for evolution in time,
while Monotonized central slope-limited linear interpolation method is
used for second-order accuracy in space~\cite{van77}. 2D scheme (2-dimensional
Newton-Raphson method) is adopted for transforming between conserved
variables and primitive variables~\cite{noble06}.
We used a simple gamma-law equation of state with $\gamma=4/3$.
Modified Kerr-Schild coordinate is adopted with mass of the black
hole ($M$) fixed where the Kerr-Schild radius $r$ is replaced by the
logarithmic radial coordinate $x_1= \ln r$. 
In the following, we use $G=M=c=1$ unit.

The computational domain is axisymmetric, with a grid that extends
from $r_{\rm in} = 0.98 r_+$ to $r_{\rm out} = 230$ and from $\theta =
0$ to $\theta = \pi$ where $r_+$ is the outer event horizon. 
The numerical resolution is 300 $\times$ 300. 
As an initial condition, we put the 0th order terms of the monopole
solution around the black hole~\cite{komissarov04}. That is, $\Re^{\mu} = - n_{\nu}
^{*}F^{\mu \nu} = (0,\alpha C \sin \theta / \sqrt{-g},0,0)$ in the
Kerr-Schild coordinate where $n_{\nu}$, $^{*}F^{\mu \nu}, \alpha, g$
are the normal observer's four-velocity, the dual field tensor,
Lapse-function, and determinant of the Kerr-Schild metric. The
numerical constant $C$ is set to be unity in this study.
The plasma velocity relative to the FIDO is set to
zero initially, and its pressure and density are set to the same value
of $P = \rho = \Re^2/100$ so that force free approximation is a good
one. Also, to keep the magnetization reasonably low, when the critical
condition $0.01 \Re^2 \ge \Gamma^2 \rho + (\gamma \Gamma^2 - (\gamma-1)) U$
is satisfied, density and internal energy $U$ is increased by the same
factor so that the critical condition holds~\cite{komissarov04}. Here
$\Gamma$ is the bulk Lorentz factor of the fluid measured in the
Kerr-Schild coordinate. We have performed numerical simulations with
the Kerr parameters 0, 0.01, 0.05, 0.1, 0.15, 0.2, 0.3, 0.4, 0.5, 0.6, 0.7, 0.8,
0.9, 0.95, 0.99, and 0.995 until time T = 200.

\begin{figure}[t]
\includegraphics[width=\linewidth]{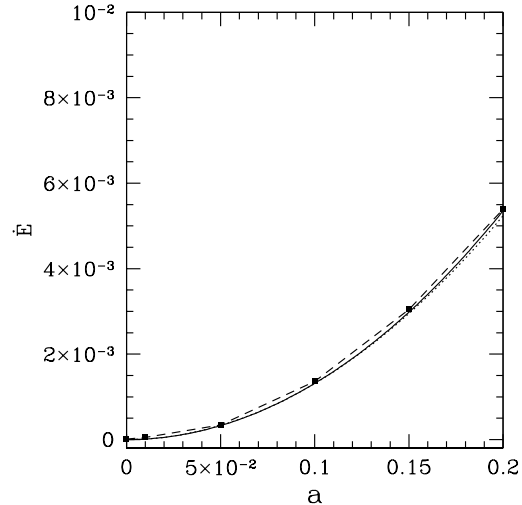}
\caption{\label{Fig1} Comparison of the derived, conserved, total energy flux.
Dashed line with rectangular points is numerical result for small Kerr
parameter ($0 \le a \le 0.2$), dotted line shows the second-order analytical
solution, and solid line represents the forth-order analytical solution.}
\end{figure}

In Fig.~\ref{Fig1}, we plot the total energy flux at the final stage
for small Kerr parameter ($0 \le a \le 0.2$) by rectangular points. 
The total energy flux was evaluated at
$r=20$, although we found that the total energy flux is insensitive to
the radius where it is evaluated. This means that the conservation of
the total energy flux has been confirmed numerically. Dashed line is just
the interpolation of the calculated values. For comparison, the
second-order analytical solution is shown by the dotted line
and the forth-order analytical solution is shown by the solid line.
From this comparison, we can see that all of them coincide with each
other that means the Blandford-Znajek solution is a really good
approximation for small Kerr parameters.

\begin{figure}[t]
\includegraphics[width=\linewidth]{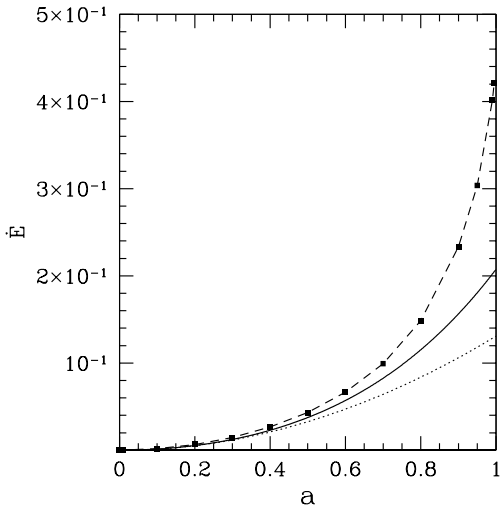}
\caption{\label{Fig2} Same with Fig.1, but for wide range of the Kerr
parameter $(0 \le a \le 1)$}
\end{figure}

The situation becomes different drastically for large Kerr parameter. 
In Fig.~\ref{Fig2}, we plot the same values with Fig.~\ref{Fig1}, but
with wide range of the Kerr parameter ($0 \le a \le 1$). We can see
clearly the difference among three cases. However, we would like to
emphasize the following point: we could confirm that the forth order
analytical solution is a better approximation rather than the
second order analytical solution. In fact, the deviation between the
numerical result and forth order analytical solution is less than that
between the numerical result and second order analytical solution.
Of course, we could also see that much higher-order analytical
solution is required to reproduce the numerical result, which is out
of the scope in this study.

\section{discussion and conclusion}
\subsection{discussion}

We have tried to solve the fourth order terms for the Blandford-Znajek
monopole solution to evaluate extracted energy flux more accurately. 
Since the equation for
$B^{\phi}$ is a first order differential equation, we could solve by
imposing a boundary condition at the horizon. However, $A_{\phi}$
obeys the second order differential equation, thus we need to put two boundary
conditions. Blandford and Znajek (1977) imposed the boundary conditions at
the horizon and infinity. The boundary condition at infinity was
chosen to connect the Michel's solution. 
In this study, we tried to impose the same
boundary conditions for the fourth order solution.
However, we found that the fourth order solution can not be connected
to Michel's solution with any boundary condition at infinity.
After all, we could not find 
the fourth order solution of $A_{\phi}$ assuming the force free condition. 
Furthermore, we found the perturbation method for Kerr parameter $a$
breaks down at large $r$ because of the behavior $A_{\phi}^{(4)} \sim
r^2$. Fortunately, the fourth order term of the total energy flux
extracted from a rotating black hole, which we want to evaluate most, 
does not depend on the solution we could not solve.


The boundary condition which we could not specify was considered
in~\cite{uzdensky04-01,uzdensky04-02}. 
In general, the equation for the stream function $A_{\phi}$ which is 
called Grad-Shafranov equation has some singular surfaces~\cite{beskin04,kim05}. 
In the force free limit, this
equation is singular at horizon and at light cylinder. Thus, in order to
determine the solution uniquely, we have to impose proper boundary
conditions at horizon and light cylinder. However, boundary conditions
to be imposed is not understood well~\cite{uzdensky04-01}.

We have compared the numerical result with our fourth order solution. The numerical 
result we use is the result in GRMHD simulation. But for the evaluation of the total 
energy flux, GRMHD result is almost same as the result in GRFFE (General Relativistic 
Force Free Electrodynamics) simulation~\cite{blandford08}, so we would get the same 
conclusion when we compare the GRFFE result with our fourth order solution. However, 
for the details of the electromagnetic fields (e.g. shape of the field
lines), the numerical result of GRMHD 
would be different from the one in GRFFE. As our future work, we want to examine the 
difference among our analytical result, the numerical result of GRMHD,
and the one of GRFFE.  

In our calculation, we assume an infinitely thin disk, and cannot self-consistently determine 
a constant $C$ in Eq.~(\ref{eq18}) which determines a magnitude of the
magnetic field.  However, in a realistic accretion disk, $C$ 
may depend on $a$~\cite{mckinney05} and total energy flux may be
determined self-consistently. Thus, 
when Blandford-Znajek mechanism is considered with a realistic accretion disk,
our result Eq.(\ref{eq:ef4}), which also contains the constant $C$, has to be treated carefully. 

\subsection{conclusion}

Our aim of this paper is to evaluate the total energy flux extracted
from a rapidly rotating black hole by Blandford-Znajek mechanism 
more accurately. As a result, although we could not the obtain the all perturbation solution up to the fourth order, 
we could evaluate the total energy flux extracted from a rotating
black hole without any ambiguity irrespective of the unknown boundary
condition at infinity. 
Also, we found the perturbation method for the fourth order terms breaks down 
at large $r$ by solving the equation for $A_{\phi}^{(4)}$ under the force free condition. 
This would be because in the monopole solutions, the force free condition can not 
be compatible with a rotating black hole at the fourth order of Kerr parameter.

From the comparison between the numerical solution that
is valid for $0<a<1$ and the fourth order solution, we find 
that the fourth order solution reproduces the numerical result better
than the second order solution.
At the same time, since the fourth order solution does not match well
with numerical result at large $a$, we conclude that more higher order
terms are required to reproduce the numerical result.

\begin{acknowledgments}
We are grateful to R. Blandford, J. McKinney, and K. Ruben for useful discussion. 
We also appreciate K. Murase and J. Aoi for important discussions and comments. 
The computation was carried out on NEC SX-8 and SGI Altix3700 BX2 at
Yukawa Institute for Theoretical Physics, Kyoto University. 
This work is in part supported by a Grant-in-Aid for the 21st Century
COE ``Center for Diversity and Universality in Physics'' from the
Ministry of Education, Culture, Sports, Science and Technology of
Japan. S.N. is partially supported by Grants-in-Aid for Scientific
Research from the Ministry of Education, Culture, Sports, Science and
Technology of Japan through No. 19047004, 19104006, and 19740139.
\end{acknowledgments}

\end{document}